\documentclass[24 pt]{article}

\begin{document}

\author{M.P John \\
International School of Photonics, \\
Cochin University of Science and Technology, \\
Kochi - 682 022, India \\
email:manupj@cusat.ac.in}
\title{Sampling with quantum mechanics }
\date{}
\maketitle

\begin{abstract}
A new algorithm for estimating the fraction of numbers that is
present in a superpositional state which satisfies a given
condition,is introduced.This algorithm is conceptually simple and
does not require quantum Fourier transform.Also the number of
steps required does not depend on the size of the data base to be
searched.
\end{abstract}
\bigskip
Quantum computation[1,2,3] offers a unique class of algorithms
based on Quantum parallelism.It has been shown that certain
problems like factorization[4] and database search[5] can be done
more efficiently in a quantum computer.In this letter I propose a
quantum algorithm for estimating the approximate number of
solutions to a given problem.

\smallskip
The given condition is
\begin{center}
$$
C_n(x)=0
$$
\end{center}
$C_n(x)$ can be a mathematical or a logical statement.Though we
are not sure what will be the final form of the quantum
computer,it can be said with some confidence that it will will
have an input register,a quantum processor,and an output register.

let the input register be represented as X and the output register be Y.The
action of the quantum processor on these registers be represented as $U_c$%
.The registers hold k qubits so that an equally weighted
superposition of $0$ to $2^k-1$ numbers can be prepared.

\smallskip

we define a new variable $y$ such that

\begin{center}
$y(x)=0$ if $C_n(x)\neq 0$

$y(x)=1$ if $C_n(x)=0$
\end{center}

Our quantum processor can calculate $y(x)$ for all values of x.An
superposition is prepared in the X register

\begin{center}
$$
2^{-k/2}\sum\limits_{j=0}^{2^k-1}\mid j\rangle
$$
\end{center}
Now perform the operation

$$
U_c\mid x\rangle _X\mid 0\rangle _Y\longrightarrow \mid x\rangle
_X\mid y(x)\rangle _Y
$$

As the input register contains both the numbers that satisfies the
condition and the numbers that does not satisfy the condition,

Thus the Y register will contain a superposition of 0 and 1.The
state of the computer after the computation can be represented as

$$
\mid Q\rangle =a\sum \mid x_s\rangle _X\mid 1\rangle _Y+b\sum \mid
x_{ns}\rangle_X \mid 0\rangle_Y
$$

Where $\{x_s\}$ are the values of x for which $C_n(x)=0$ and $\{$
$ x_{ns}\},$ the values of x for which $C_n(x)\neq 0$

\smallskip

The coefficients a and b depends on the number of elements in the set $%
\{x_s\} $.and $\{x_{ns}\}$that is $\mid a\mid ^2$will be
proportional to the number of elements in the set $\{x_s\}$ and
$\mid b\mid ^2$will be proportional to the number of elements in
the set $\{x_{ns}\}$.It can be seen from the above equation that
when measured,the y register projects to the
state $\mid 1\rangle _Y$ with a probability $\mid a\mid ^2$and to the state $%
\mid 0\rangle _Y$ with a probability $\mid b\mid ^2$.So after
making P
measurements of the $Y$ register one can expect $\mid a\mid ^2P$ ones and $%
\mid b\mid ^2P$ zeros.Thus by repeating the experiment $P$ times
an estimate of $\mid a\mid^2$and$\mid b\mid^2$ can be made.The
fraction of numbers f, satisfying the condition $C_n(x)=0$ can be
estimated from this.\\

It is to be noted that the number of steps required depends only
on the accuracy to which the f has to be estimated.And is
completely independent of the size of the data base,that has to be
considered.This is a remarkable feature as far as quantum
algorithms of practical use are concerned.This method does not
require the Fourier transforming or phase estimation techniques
used in the existing algorithm for quantum counting[6] and also is
conceptually much simpler.At first sight it may seem that this can
be done with classical randomness. i.e. simply select an element x
uniformly at random and test if $C_n(x)=0$.This works only if the
solutions are uniformly distributed among the non solutions.In
actual situations,the solutions may be non uniformly distributed
and one has to use statistical techniques for sampling in which
the size of the data base has to be considered at least
indirectly.Here every entry of the data-base contributes to the
final amplitude.\\

I am grateful to Prof M.Sabir, Dept of Physics, Cochin University
of Science and Technology, for giving necessary guidance in the
initial stages of the work. I am also grateful to my colleagues
and all those who have given me encouragement and helpful
suggestions.The author is supported by the Council for scientific
and Industrial Research (CSIR),India

\end{document}